\documentclass[sigconf]{acmart}
\usepackage{enumitem}
\usepackage{amsmath}
\usepackage{graphicx}

\AtBeginDocument{
	\providecommand\BibTeX{{%
			\normalfont B\kern-0.5em{\scshape i\kern-0.25em b}\kern-0.8em\TeX}}}

\usepackage{multirow}
\usepackage{soul}
\usepackage{mathtools}
\usepackage{amsmath,amsfonts, amsthm}
\usepackage{graphicx}
\usepackage{float}
\usepackage{algorithmic}

\usepackage{lipsum}
\usepackage{subfig}
\usepackage{textcomp}
\usepackage{xcolor}
\usepackage{placeins}
\usepackage{hhline}
\usepackage{tabularx}
\usepackage{enumitem}
\usepackage{setspace}
\usepackage{booktabs}
\usepackage{adjustbox}
\usepackage[normalem]{ulem}
\usepackage{indentfirst}
\usepackage{makecell}
\usepackage{titlesec}
\usepackage[T1]{fontenc}
\usepackage{textcomp, libertine}

\makeatletter
\newcommand{\mathleft}{\@fleqntrue\@mathmargin0pt}
\newcommand{\mathcenter}{\@fleqnfalse}
\makeatother

\theoremstyle{definition}

\let\sigproof\proof\let\proof\relax
\let\sigendproof\endproof\let\endproof\relax
\let\proof\sigproof
\let\endproof\sigendproof

\DeclareMathOperator*{\argmax}{arg\,max}

\makeatletter
\def\@copyrightspace{\relax}
\makeatother

\author{Zeeshan Memon}
\affiliation{%
  \institution{Emory University}
  \city{Atlanta}
  \country{USA}}
\email{zeeshan.memon@emory.edu}

\author{Chen Ling}
\affiliation{%
  \institution{Emory University}
  \city{Atlanta}
  \country{USA}}
\email{chen.ling@emory.edu}

\author{Ruochen Kong}
\affiliation{%
  \institution{Emory University}
  \city{Atlanta}
  \country{USA}}
\email{ruochen.kong@emory.edu}

\author{Vishwanath Seshagiri}
\affiliation{%
  \institution{Emory University}
  \city{Atlanta}
  \country{USA}}
\email{vishwanath.seshagiri@emory.edu}

\author{Andreas Zufle}
\affiliation{%
  \institution{Emory University}
  \city{Atlanta}
  \country{USA}}
\email{azufle@emory.edu}

\author{Liang Zhao}
\affiliation{%
  \institution{Emory University}
  \city{Atlanta}
  \country{USA}}
\email{liang.zhao@emory.edu}

\begin{document}
\title{Deep Identification of Propagation Trees in Graph Diffusion}


	
\begin{abstract}
Understanding propagation structures in graph diffusion processes, such as epidemic spread or misinformation diffusion, is a fundamental yet challenging problem. While existing methods primarily focus on source localization, they cannot reconstruct the underlying propagation trees—i.e., "who infected whom", which are substantial for tracking the propagation pathways and investigate diffusion mechanisms. In this work, we propose Deep Identification of Propagation Trees (DIPT), a probabilistic framework that infers propagation trees from observed diffused states. DIPT models local influence strengths between nodes and leverages an alternating optimization strategy to jointly learn the diffusion mechanism and reconstruct the propagation structure. Extensive experiments on five real-world datasets demonstrate the effectiveness of DIPT in accurately reconstructing propagation trees.
\end{abstract}

\keywords{Graph Source Localization, Information Diffusion, Propagation Path}

\maketitle

\section{Introduction}\label{sec: Intro}
\begin{figure}[t] 
    \centering
    \includegraphics[width=\columnwidth]{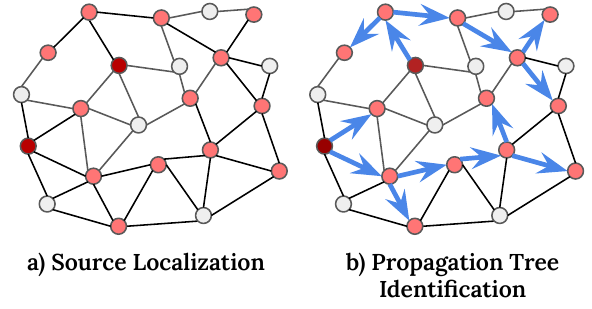} %
    \caption{Given the diffused state (colored nodes), source localization aims to identify the source nodes (red) only (a), while propagation trees identification can further reveal how infection spreads, as shown in blue arrows (b), beyond just source identification.}
    \label{fig:intro}
\end{figure}

Graph inverse problems aim to uncover the underlying causes of observed phenomena in networks. A key focus in recent research is diffusion source localization, which seeks to identify the origins of information diffusion within a network. This problem has attracted significant interest in algorithm design, and with the success of deep graph learning across various domains, it has also been extended to source localization~\cite{ling2022source,yan2024diffusion,wang2023lightweight,xu2024pgsl,ling2024source}. However, source localization alone does not provide sufficient insight into the underlying diffusion process that led to the observed spread. As illustrated in Fig.~\ref{fig:intro}, source localization (a) identifies only the source nodes from all infected nodes(pink nodes), whereas propagation trees identification (b) captures the full dynamics of diffusion, revealing not only the sources but also how the infection spread and who infected whom.

Understanding the pathways of information diffusion in a graph is an abstract problem but has broad impacts in many specific applications: Applied to infection diseases ecology, this problem of understanding who infected whom is also known as Contact Tracing~\cite{mokbel2020contact,kleinman2020digital}, and used to warn, quarantine, or hospitalize individuals who may have been exposed to an infectious disease. 
In Misinformation Spread, understanding who shared misinformation with whom may help us understand users of a social network who deliberately spread misinformation, and take action to reduce this spread or help fact-checking~\cite{zhou2019network}.
In Phylogenetics, accurately tracing the pathways of genetic mutations helps to build evolutionary trees and understand the spread of genetic traits, which is fundamental for understanding species evolution~\cite{penny2004inferring,bouckaert2014beast}.

 The problem of propagation tree identification is defined as the process of reconstructing the path from source nodes to diffused nodes using only the observed diffusion data. Despite its importance, this task remains largely under-explored as learning to reconstruct propagation paths in inverse problems presents several challenges:
    \textit{1) Necessity and Difficulty to Infer the unknown propagation mechanism.}  Existing methods rely on traditional predefined models such as Linear Threshold (LT), Information Cascading (IC), and Susceptible-Infectious-Susceptible (SIS), which make simplified and rigid assumptions. However, these models fail to capture the complexity of real-world scenarios, where infection dynamics depend on multiple node attributes. For example, whether a rumor will be passed from Users A to B depends on their education level, topic of interest, culture, etc. Data driven methods like graph deep learning require the observation of "who infects who", which is prohibitively difficult to be jointly learned with propagation mechanism. 
    \textit{2) Large Search Space under forest-structured constraints. }The number of possible propagation trees grows exponentially with the number of nodes and edges, making direct optimization highly intractable. 
    \textit{3) No or incomplete observation of propagation tree during training.} In many real-world scenarios, only a partial observation of the diffusion process at edge level is available. For example, in epidemiological modeling, contact tracing rarely provides the full transmission path, leading to uncertainty in propagation tree reconstruction~\cite{ rodriguez2014uncovering}. The absence of complete propagation tree data makes it challenging to learn the tree without directly observing it during training.

In this work, we propose a probabilistic framework, Deep Identification of Propagation Trees (DIPT), to identify propagation trees given diffused observations. To address Challenge~1, we model the diffusion process recursively, where each node's infection probability is learned based on its parent nodes' infection states and their features. This allows the propagation mechanism to dynamically adapt to the underlying structure of the data, without the need for rigid assumptions.
For Challenge~2, we constrain the exponential search space by leveraging prior information over source nodes, which limits candidate trees to those originating from likely sources, and then infer the most probable propagation tree that maximizes the likelihood of the observed diffusion states.
To address Challenge~3, where propagation trees' data is not available during training, we use an alternating optimization strategy that jointly learns the propagation trees and propagation mechanism.
We summarize the major contributions of this work as follows:
\begin{itemize}[left=0pt]
    \item \textbf{New Problem} We highlight novel problem in graph inverse problems, as generalization of source localization where instead of sources only we also identify propagation tree explaining underlying diffusion pattern.
    \item \textbf{Training Algorithm} We propose alternating optimization training algorithm to learn propagation tree and diffusion model without having propagation tree as observable during training.
    \item \textbf{Modeling and Inference} We model a unified objective to learn diffusion pattern, by learning local propagation influence of nodes using node attributes, for identifying the diffusion sources and propagation tree.
    \item \textbf{Extensive Experiments.} We conduct experiments against state-of-the-art methods and results show consistent performance of our method for path reconstruction and source localization against five datasets.
\end{itemize}

\section{Related Work}\label{sec: LitReview}
\noindent\textbf{Source Localization.}
Diffusion source localization, the task of identifying the origins of diffusion from observed spread patterns, has applications ranging from rumor detection in social networks to tracing blackout sources in smart grids ~\cite{jiang2016identifying, shelke2019source}. Early studies \cite{prakash2012spotting, wang2017multiple, zhu2016identifying, zhu2016information, zang2015locating, zhu2017catch} concentrated on identifying single or multiple sources within classical diffusion frameworks, such as Susceptible-Infected (SI) and Susceptible-Infected-Recover (SIR), leveraging either complete or partial observations. More recently, advancements like the work by Dong et al.\cite{dong2019multiple} employed Graph Neural Networks (GNNs) to enhance prediction accuracy. However, many existing methods face challenges in quantifying the uncertainty among candidate sources and require computationally expensive searches over high-dimensional graph structures, which limits their scalability and practicality. These limitations make it difficult to effectively apply these methods in large-scale, real-world scenarios. In response, newer approaches ~\cite{ling2022source, qiu2023reconstructing, wang2022invertible, wang2023lightweight, xu2024pgsl} have shifted focus toward mitigating the dependency on predefined diffusion models and characterizing latent source distributions, achieving state-of-the-art results. Nevertheless, these methods cannot handle the problem of reconstructing diffusion paths, which remains a critical challenge.
\newline \newline
\textbf{Propagation Path Reconstruction}. Traditional approaches leverage historical interaction data to quantify link weights (e.g., tie strength) and employ algorithms like LeaderRank for ranking influential nodes, followed by evaluating path importance through probabilistic measures~\cite{zhu2016identifying}. Advanced deep learning-based models, such as I3T, integrate graph neural networks (GNNs) for local neighborhood aggregation, sequence sampling, and Bi-LSTM and GRU architectures to extract structural and temporal diffusion features, enhanced with attention mechanisms~\cite{tai2023predicting, xia2021deepis}. Additionally, fast prediction techniques and topic-oriented relationship strength networks provide efficient path prediction while capturing community-level diffusion dynamics~\cite{zhu2023path}. These methodologies focus on the propagation path in forward information diffusion tasks. Recent approaches have shifted towards leveraging deep generative models to capture more complex diffusion dynamics. Two recent methods leverage deep generative models to construct propagation path cascades. DDMIX \cite{vcutura2021deep} uses a Variational Autoencoder (VAE) to learn node states across all time steps and reconstruct dissemination paths. However, as the number of time steps grows, the increasing solution space complexity reduces its accuracy in source localization. Building on this, Discrete Diffusion Model for Source Localization (DDMSL) employs deep generative diffusion models to reconstruct the evolution of information diffusion and perform source localization\cite{yan2024diffusion}. While both methods can generate snapshots of diffusion—identifying infected nodes at specific times—they cannot capture the complete propagation path i.e. the exact transmission edges between nodes.
\newline \newline
\textbf{Alternating Optimization in Graph Problems.} Alternating optimization techniques are widely used in graph-based learning problems to iteratively refine model parameters and latent or unobservable variables. For instance, Learning to Propagate~\cite{xiao2021learning} employs an alternating update strategy within a Graph Neural Network (GNN) to learn personalized propagation strategies for nodes. GLEM~\cite{zhao2022learning} optimizes language models and GNNs in an alternating manner to efficiently learn on large text-attributed graphs. A hybrid MLCO approach~\cite{wang2021bi} applies bi-level optimization for graph structure learning and combinatorial problem-solving. These methods demonstrate the effectiveness of alternating optimization in handling unobserved variables and improving inference in graph-based models. Motivated by this, our approach incorporates a similar alternating strategy to model propagation paths in graph diffusion process.

\section{Deep Identification of Propagation Trees}\label{sec: model}
In this section, we first present the problem formulation and define a unified objective function to learn the diffusion pattern based on node features. Subsequently, we propose an alternating optimization method to jointly learn the propagation path and diffusion pattern.
    
\subsection{Problem Formulation}\label{sec: formulation}
Given a graph \( G = (V, E, \mathbf{y}) \), where \( V \) is the set of nodes, \( E \subseteq V \times V \) is the set of edges, and \( \mathbf{y} \in \{0,1\}^{|V|} \) is a binary infection state vector indicating whether node \(v\) is infected (\( \mathbf{y}_v = 1 \)) or not (\( \mathbf{y}_v = 0 \)), information spreads through propagation trees \(\mathcal{T}\) from source nodes \( \mathbf{s} \in \{0,1\}^{|V|} \), \( \mathbf{s}_v = 1 \) denotes that node \( v \) is a diffusion source, and \( \mathbf{s}_v = 0 \) otherwise. Specifically, \(\{T_v\}_{s_v=1}\), where \({T_v}\) is a propagation tree rooted at seed node \(v\). Each node is associated with a feature vector of dimension \( F \), forming a feature matrix \( \mathbf{F} \in \mathbb{R}^{|V| \times F} \).

The objective is to infer the propagation trees \(\mathcal{T}\) given the observed infection state \(y\). This task faces the following significant challenges:\newline
\textbf{Challenge~1: Unknown Propagation Mechanism.} The diffusion process is influenced by node attributes $\mathbf{F}$, such as age groups or shared interests among users. These attributes increase the likelihood of mutual influence between similar nodes. Existing methods, which rely on blackbox deep learning models, struggle to accurately learn diffusion patterns and reconstruct propagation trees as they fail to account for these influential node attributes.\newline
\textbf{Challenge~2: Large Search Space for Propagation Paths.} The number of possible propagation trees grows exponentially with the graph size, making direct optimization over all possible propagation paths computationally infeasible.\newline
\textbf{Challenge~3: Incomplete Observation of Propagation Trees.} In real-world scenarios, only partial diffusion data is available at the edge level. For example, in epidemiology, contact tracing often provides incomplete transmission paths, making it difficult to learn the propagation tree without observing it during training.

\subsection{Objective Function}
The objective is to find the optimal source nodes \( \tilde{s} \) and propagation path \( \tilde{\mathcal{T}} \) that maximizes joint probability~\( P(s, \mathbf{y} \mid \mathcal{T}, G) \), we utilize the graph topology \( G \) and diffusion observations to define the joint probability: 
\begin{equation}
\tilde{s}, \tilde{\mathcal{T}} = \argmax_{s, \mathcal{T}} P(s, \mathbf{y} \mid \mathcal{T}, G).    
\label{eq:objective_1}
\end{equation}  
However, Eq.~\eqref{eq:objective_1} cannot be solved directly. Since the infection state \( y \) is influenced by the graph topology \( G \), source nodes, \( s \), and the propagation tree \( \mathcal{T} \), we can decompose the problem. This reformulation simplifies the Maximum A Posteriori (MAP) estimation as follows:
\begin{equation}
P(s, \mathbf{y} \mid \mathcal{T}, G) = P(\mathbf{y} \mid s, \mathcal{T}, G) \cdot P(s),
\label{eq:MAP_simplified}
\end{equation}  
where:  
\begin{itemize}[left=0pt]
\item \( P(s) \): The prior distribution over the source nodes.  
\item \( P(\mathbf{y} \mid s, \mathcal{T}, G) \): The likelihood of the observed infection state \( \mathbf{y} \), given the source nodes \( s \) and the propagation tree \( \mathcal{T} \) in \( G \).  
\end{itemize}
From Eq.~\eqref{eq:MAP_simplified}, we parametrize prior of seed nodes and likelihood of observed infection states with \(\phi\) and \(\psi\) respectively. Therefore,  based on Eq.~\eqref{eq:MAP_simplified}, Eq. \eqref{eq:objective_1} can be rewritten as:
\begin{equation}
\max_{\mathcal{T}, \phi, \psi} P_{\psi}(y | s, \mathcal{T}, G) \cdot P_{\phi}(s)    
\label{eq:objective}
\end{equation}
We learn \(\phi, \psi\) and \(\mathcal{T}\) in training phase. Further explanations are provided in subsequent subsections.

\subsection{Estimation of the Graph Diffusion Process}
\label{sec:Diffusion}
This section introduces the solving of \(\psi\) with respect to Eq.~\eqref{eq:objective}.
The infection probability of any node \( y_n \) is determined by the infection states of its parent nodes, denoted as \( \operatorname{Pa}(y_n, \mathcal{T}) \). Unlike modeling diffusion process as direct mapping from seed to diffused nodes as in prior work, we model the joint probability of the diffusion process as a product of conditional probabilities, recursively expanding over all affected nodes:  
\begin{equation}
P_{\psi}(y \mid s, \mathcal{T}, G) = \prod_{n \in V_{\setminus s}} P_{\psi}(y_n \mid \operatorname{Pa}(y_n, \mathcal{T})).
\label{eq:06}
\end{equation}  
where \( V _{\setminus s} \) denotes all nodes excluding the sources.    
The conditional probability \( P(y_n \mid \operatorname{Pa}(y_n, \mathcal{T})) \) represents the influence of a node on another along an edge and is parameterized by a learnable model that computes the infection probability of \( y_n \) based on the features of both \( y_n \) and its parent node \( \operatorname{Pa}(y_n, \mathcal{T}) \). Formally, \( f_\psi \) is defined as:  
\begin{equation}   
P_{\psi}(y_n \mid \operatorname{Pa}(y_n, \mathcal{T})) = f_\psi(\mathbf{F}_{y_n}, \mathbf{F}_{\operatorname{Pa}(y_n)}),
\label{eq:07}
\end{equation}  
where \( \mathbf{F}_{y_n} \) and \( \mathbf{F}_{\operatorname{Pa}(y_n)} \) are the feature vectors of \( y_n \) and its parent node, respectively. That means, Challenge~1 can be addressed by recursively modeling the diffusion process based on node features. The parameters \(\psi\) are optimized to maximize the likelihood of observed infections, ensuring that the predicted infection states of nodes align with the observed infected data. This alignment enables the model to effectively learn the underlying diffusion pattern.
\begin{align}
\mathcal{L}_{\text{Diffusion}} = 
&\nonumber -\sum_{n \in V} \Big[
    y_n \log {P}_\psi\left(y_n \mid \operatorname{Pa}(y_n, \mathcal{T})\right) \\
    & + (1 - y_n) \log \left(1 - {P}_\psi\left(y_n \mid \operatorname{Pa}(y_n, \mathcal{T})\right)\right)
\Big].
\label{eq:08}
\end{align}


\subsection{Optimizing Propagation Trees}\label{LEarning}
In this section, we infer the most probable propagation tree \( \mathcal{T}^* \) by optimizing the objective function in Eq.~\eqref{eq:objective}. Since the true propagation tree is unobserved during training, we estimate \( \mathcal{T}^* \) by maximizing the joint likelihood while keeping the model parameters fixed with respect to \(\mathcal{T}\). The inferred tree consists of directed edges linking seed nodes \( s \) to observed infected nodes \( y \).:  
\begin{equation}
\mathcal{T}^* = \argmax_{\mathcal{T}} P_{\psi}(y \mid s, \mathcal{T}, G) \cdot P_{\phi}(s).
\end{equation}  

As explained in Section~\ref{sec:Diffusion}, infection spreads iteratively. At the initial step, \( k = 0 \), the infection begins at the seed nodes \( s \), with their infection probabilities given by \( P_{\phi}(s) \). Infection then propagates along the edges \( (i, j) \) of the graph, where the spread is determined by the influence scores between nodes. These influence scores are parameterized by a learnable function \( f_\psi \), as described in Eq.~\eqref{eq:07}, and the resulting scores are stored in the influence matrix \( \mathbf{I} \). Each entry \( \mathbf{I}_{(i, j)} \) represents the probability that node \( i \) can infect node \( j \), as learned by the parameterized function \( f_\psi \).
\begin{equation}
P_{\text{infected}}(0) = P(s) \cdot I.
\end{equation}  
where \( P_{\text{infected}}(0) \in \mathbb{R}^{|V|} \) is a vector representing the infection probabilities for each node in the graph. At each iteration \( k+1 \), infection probabilities are updated as:  
\begin{equation}
P_{\text{infected}}(k+1) = P_{\text{infected}}(k) \cdot I.
\end{equation}  
This process continues until infection reaches the target nodes \( y \). However, not all nodes update at every step. A mask is applied to selectively update only those nodes whose infection probabilities increase compared to the previous iteration, ensuring that the inferred propagation path maximizes the likelihood of observed infections.
The mask for node \( i \) at iteration \( k \) is defined as:  
\begin{equation}
\mathbf{M}_k(i) = \left( (P_{\text{infected}}(k-1) \cdot \mathbf{I})_i > P_{\text{infected}}(k-1)_i \right)
\end{equation}
where \( P_{\text{infected}}(k-1)_i \) is the infection probability for node \( i \) at iteration \( k-1 \).

The binary mask \( \mathbf{M}_k(i) \in \{0, 1\} \) ensures that only nodes with increased infection probability are updated. This can be expressed as:
\begin{equation}
P_{\text{infected}}(k) = \left( P_{\text{infected}}(k-1) \cdot \mathbf{I} \right) \odot \mathbf{M}_k
\end{equation}
where, \( \odot \) denotes the element-wise product.
Once infection probabilities are computed for all nodes, we construct the most probable propagation tree \( \mathcal{T}^* \) by selecting edges that maximize the likelihood of infection transmission. Specifically, for each infected node \( i \), we determine its parent \( p_i \) as the node that maximizes the conditional infection probability:  
\begin{equation}
p_i = \argmax_{j \in \mathcal{C}(i)} P(i \mid j).
\end{equation}  
where \( \mathcal{C}(i) \) denotes the set of neighboring infected nodes for \( i \). The resulting propagation trees are then the set of these selected edges:  
\begin{equation}
\mathcal{T}^* = \{(p_i, i) \mid i \in y \}.
\end{equation}  
Since each node selects exactly one parent that maximally contributes to its infection probability, \( \mathcal{T}^* \) forms directed trees rooted at the seed nodes, spanning all infected nodes.  

By following this approach, we infer the most probable propagation structure, ensuring that \( \mathcal{T}^* \) represents the most likely sequence of transmissions from the seed nodes to the observed infections.

\subsection{Learning the Prior of Seed Nodes}

The intrinsic patterns of the prior over seed nodes, $P(s)$, are hard to model and often high dimensional, which leads to intractability. To tackle it, we propose to map $s$ to a latent embedding $z$ residing in lower dimensional space representing the abstract semantics. A variational inference framework is used to learn an approximation of \( P(s) \), capturing its structure and variability. It consists of a generative process \( P_{\phi_2}(s \mid z) \) that reconstructs \( s \) from \( z \), and a simple prior \( P(z) \), typically a standard Gaussian \( \mathcal{N}(0, I) \), which regularizes the latent space. The variational posterior \( q_{\phi_1}(z \mid s) \) approximates the intractable \( P(z \mid s) \) and serves as the encoder. Variational inference is introduced to efficiently approximate the intractable posterior \( P(z \mid s) \) by optimizing the Evidence Lower Bound (ELBO), ensuring that the latent variables \( z \) capture the meaningful variability of \( s \) while maintaining regularization through the prior \( P(z) \). The objective is to maximize the Evidence Lower Bound for \( P(s) \):  
\begin{equation}
L_{\text{ELBO}} = \mathbb{E}_{q_{\phi_1}(z \mid s)} \left[\log P_{\phi_2}(s \mid z)\right] - \text{KL}\left(q_{\phi_1}(z \mid s) \| P(z)\right).
\label{eq:elbo}
\end{equation}  
The first term ensures accurate reconstruction of \( s \), while the second term regularizes the latent distribution to match the prior \( P(z) \), using the Kullback-Leibler~(KL) divergence as a measure of divergence between the variational posterior and the prior. This formulation ensures that \( z \) captures meaningful variations in \( s \) while maintaining a structured representation.  

\subsection{Incorporating Partial Propagation Path Observations}
\label{sec:partial}
The propagation tree is unobservable during training, so the diffusion process is inferred solely from the observed infection states \( \mathbf{y} \). However, when partial information about the propagation tree, \( \mathcal{T}_{\text{obs}} \), is available, a supervised component is added using the observed edges. This helps the model learn the propagation tree structure more effectively. The objective is then revised to maximize the joint probability while incorporating the observed edges: 
\begin{equation}
\begin{aligned}
P(s, \mathbf{y} \mid \mathcal{T}_{\text{obs}}, \mathcal{T}_{\text{unobs}},  G) = \,
& P(\mathbf{y} \mid s, \mathcal{T}_{\text{obs}} \cup \mathcal{T}_{\text{unobs}}, G) \cdot P(\mathcal{T}_{\text{obs}} \mid s, y, G)\\\cdot P(s)
\end{aligned}
\label{eq:joint_prob}
\end{equation}

Here, \( P(\mathcal{T}_{\text{obs}} \mid s, y, G) \) represents the likelihood of observed edges, while \( {T}_{\text{unobs}}\) still needs to be inferred the same way as before. The observed edges provide direct supervision, aiding the model in estimating the unobserved diffusion paths.

This inclusion of observed edges \( \mathcal{T}_{\text{obs}} \) allows direct optimization over the edges of the observed tree. For each observed edge \( (u, v) \in \mathcal{T}_{\text{obs}} \), we maximize the conditional probability \( P_\psi(Y_u \mid Y_v) \), representing the likelihood of node \( u \) being infected given node \( v \) is infected. This introduces a supervised loss term:
\begin{equation}
\label{eq:supervisedEdge}
\mathcal{L}_{\text{observed}} = - \sum_{(u, v) \in \mathcal{T}_{\text{obs}}} \log P_\psi(y_u \mid y_v).    
\end{equation}

Thus, the total objective function in Equation~\ref{eq:objective} can be written as:  
\[
\begin{aligned}
\min_{\theta, \phi, \psi} \mathcal{L}_{\text{train}}(\theta, \phi, \psi) = & \mathbb{E}_{q_\phi(Z \mid s)} \left[ -\log P_\theta(s \mid z) \right] \\
& + \text{KL}\left[ q_\phi(z \mid s) \| P(z) \right] \\
& + \lambda \cdot \left( -\sum_{n \in V} \left[ 
y_n \log \hat{P}_\psi(y_n \mid \operatorname{Pa}(y_n, \mathcal{T})) \right. \right. \\
& \quad \left. \left. + (1 - y_n) \log \left(1 - \hat{P}_\psi(y_n \mid \operatorname{Pa}(y_n, \mathcal{T}))\right) 
\right] \right) \\
& + \mu \cdot \left( - \sum_{(u, v) \in \mathcal{T}_{\text{obs}}} \log P_\psi(y_u \mid y_v) \right),
\end{aligned}
\]  
where \( \lambda \) and \( \mu \) are hyperparameters controlling the relative contributions of the diffusion and observed edges loss terms, respectively. Thus is case of partially observed propagation tree, the model transitions from being entirely unsupervised, where it infers propagation tree without any ground truth about the propagation path, to a partially supervised framework.

\subsection{Propagation Tree Inference}
\label{sec:propTreeInference}
During inference, we aim to identify optimal propagation tree rooted at seed nodes by initializing a learnable latent vector \( \hat{z} \), which is iteratively optimized while keeping all model parameters (\( \phi \), and \( \psi \)) fixed, which means instead of maximizing joint probability, shown in Eq.~\ref{eq:objective}, with respect to model parameters we now maximize with respect to \(z\). To provide a starting point from the training distribution, \( \hat{z} \) is initialized to the average latent representation \( \bar{z} \), computed as the mean of the encoded latent vectors from the training dataset. Similar to learning framework, optimum propagation tree \(\mathcal{T^*}\) is estimated in the same way as in training stage. The optimization problem can be defined as:
\begin{equation}
\mathcal{L}_{\text{inf}} = \max_{\hat{z}} \mathbb{E}\left[{p_\psi(y \mid s, \mathcal{T}^*, G)} \cdot p_{\phi_{2}}(s \mid \hat{z}) \right],
\label{eq:inf}
\end{equation}
To ensure that the learnable latent vector \( \hat{z} \) does not diverge excessively during inference, an additional regularization term is introduced. This term constrains \( \hat{z} \) to remain close to the mean of the latent distribution \( \bar{z} \). The total loss function during inference becomes:
\begin{equation}
\mathcal{L}_{\text{inf}} = \max_{\hat{z}} \mathbb{E}\left[ p_\psi(y \mid s, \mathcal{T}^* G) \cdot p_\theta(s \mid \hat{z}) \right] - \gamma \cdot \| \hat{z} - \bar{z} \|^2
\label{eq: infer}
\end{equation}
The constrained objective function Eq.~\eqref{eq: infer} cannot be computed directly, so we provide a practical version of the inference objective function: since the diffused observation \( y \) fits the Gaussian distribution and the seed set \( s \) fits the Bernoulli distribution, we can simplify Eq. (7) as:  
\begin{equation}
\begin{aligned}
\mathcal{L}_{\text{inf}}^* &= \min_{\hat{z}} \Bigg[ -\sum_{i=1}^{N} \log \left( f_{\phi_{2}}(\hat{z})^{s_i} \cdot (1 - f_{\phi_{2}}(\hat{z}))^{1 - s_i} \right) \\
& \quad + \frac{1}{2} \| \hat{y} - y \|^2 + \lambda \| \hat{z} - \bar{z} \|^2 \Bigg]
\end{aligned}
\end{equation}
Here, the first term captures the Bernoulli likelihood of the seed nodes \( s \) and \(f_{\phi_{2}}\) denotes the decoder in VAE, while the second term represents the Gaussian likelihood by minimizing the squared error between the predicted diffused observation \( \hat{y} \) and the true observation \( {y} \).

\section{Experimental Evaluation}
The proposed method, DIPT, is evaluated on five real-world datasets to address the following research questions:

\begin{itemize}[leftmargin=*]
    \item \textit{How accurately does DIPT identify the propagation path} compared to existing methods?
    \item \textit{How effectively does DIPT identify diffusion sources} compared to baseline methods?
    \item \textit{How well does DIPT reconstruct node states} during the diffusion process?
    \item \textit{What is the impact of incorporating partial propagation tree observations} during training on DIPT's performance?
    \item \textit{How does each component of DIPT contribute to the overall system?}
\end{itemize}
\begin{figure*}[t] 
    \centering
    \includegraphics[width=1.00\linewidth]{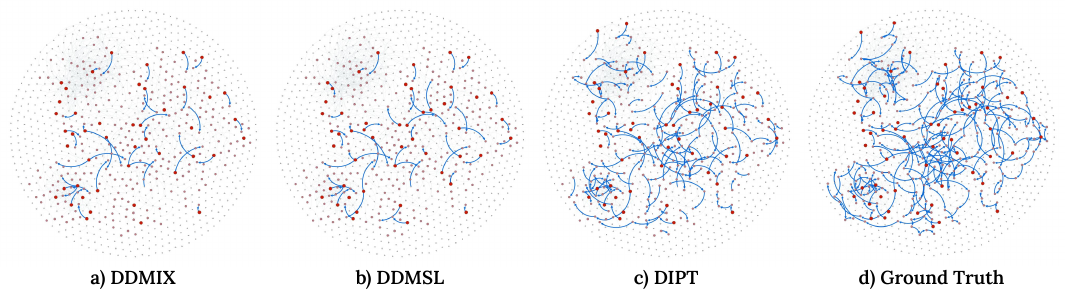} %
    \caption{Comparison of correctly predicted propagation tree edges (blue) with ground truth for the MemeTracker dataset. Source nodes are in red, infected nodes in pink. Only correctly predicted edges are shown for clarity, with the total number of predicted edges being the same across all methods.}
    \label{fig:comparison_memetracker}
\end{figure*}
\subsection{Datasets}
\subsubsection{\textbf{\large {Real-World Datasets.}}}
We evaluate the proposed model, DIPT, with baseline methods using five real-world datasets. 
For the Cora-ML~\cite{cora}, CiteSeer~\cite{wu2016citeseerx}, and Power Grid~\cite{watts1998collective} datasets, information diffusion data is not available; however, these datasets provide graph topology and node features. To simulate diffusion, we randomly select 10\% of the nodes as source nodes, and the spread of information is then modeled using the SI epidemic model for 200 iterations until convergence. In contrast, the MemeTracker dataset explicitly captures real-world information propagation through hyperlinks between online articles, forming diffusion cascades. Each cascade represents how information spreads across different sites over time, as described in~\cite{rodriguez2011uncovering}. For our experiments, we extract a subnetwork consisting of the top 583 sites and 6,700 cascades. Within each cascade, we identify the top 5\% of nodes as source nodes, selecting them based on their earliest appearance in the cascade. Each cascade is then represented as a propagation tree, providing a more direct representation of the diffusion process.

\subsubsection{\textbf{\large{Simulated Dataset: Infectious Disease Spread}}}~\\
\label{sec:Simulated}
We implemented a large-scale spatial compartmental Susceptible-Infectious-Recovered (SIR) model~\cite{kermack1932contributions} using fine-grained human mobility patterns~\cite{kang2020multiscale} to simulate the spread of infectious disease in the United States. We refer to this simulated dataset as the Infectious Disease Spread Simulation (IDSS) dataset.
In this simulation, each of the 3,143 counties of the United States is represented as a population of susceptible (S), infectious (I), and recovered (R) individuals. For each Each County $C$ the susceptible population $S_C$ is initially set to the counties population matching the U.S Census data~\cite{census}. The corresponding infectious population $I_C$ and recovered population $R_C$ of County $C$ are initialize are $I_C=R_C=0$.

Whenever an individual, in any county, is in the (I) state, there is a probability of $P_i$, called the time-dependent reproduction number, of infecting another individual where $i$ corresponds to the number of days since the individual has become infected. The $\sum_i P_i$ is the expected number of infections per agent, also known as the basic reproduction number $R_0$~\cite{delamater2019complexity}.

We model the spread of an infectious disease using the mobility flow data published in~\cite{kang2020multiscale} which captures, for each pair of counties $(A,B)$ in the United States, the estimated flow (in number of individuals) from County $A$ to County $B$ estimated using SafeGraph Foot Traffic Data~\cite{safegraph}. 
When an individual in County $A$ infects another individual, then this mobility model is used to determine the County $B$ of the newly infected individual. 
This is done by selecting a County $X$ chosen randomly weighted by the mobility out-flows of County $A$, and then selecting a County $B$ chosen randomly by the mobility in-flows of County $X$. The intuition here is to find an individual (in County $B$) that visited the same county as the already infected individual (in County $A$). We note that it is possible, and often likely, to have $A=B$, thus having the new infection in the same county as the old infection. Then, for County $B$ the variable $S$ is decremented and the variable $I$ is incremented. After $n$ days, where $n$ is the infectious period of the simulated disease, the individual is no longer infectious and the variable $I$ of the corresponding county is decremented and the variable $R$ is incremented. 
To create infectious disease data used in the following experiments, we run this simulation having for an infectious period $n=6$ days with a time-dependent reproduction number of $P=[0.2,0.3,0.3,0.2,0.1,0.1]$ having $\sum_i P_i=1.2=:R_0$ to simulate an infectious disease having pandemic potential~\cite{delamater2019complexity}. 

To start the simulation with initial cases, we simulate the arrival of an airplane with infected passengers. We use the counties in the United States having the 72 largest airports. For each simulation run, we select two of these counties as the initial sources. We then infect ten initial individuals in counties chosen randomly weighted by the mobility out-flows of these two counties. The intuition of this approach is to simulate the arrival of infected individuals by plane and going to their home counties. 

We run the simulation for $90$ simulation days, yielding approximately 500-1000 infected counties having approximately 12k-48k infected individuals per simulation run. 
For each simulation run, we report the infection forest. In this forest, each node is an infected individual and edges denote who-infected-whom relationships. Since each simulation has multiple initial infections, the resulting graph is a forest, where each node is either isolated or has exactly one path to one of the initially infected individuals, which are the roots of their trees. 

The simulation used to generate this dataset, documentation, as well as the simulated datasets for this evaluation can be found at~\url{https://anonymous.4open.science/r/Compartmental-Infectious-Disease-Simulation-4F7C}.

\subsection{Experiment Setup}
\subsubsection{\textbf{Comparison Methods}}
\label{sec:comparMethod}
\begin{table*}[!h]
\centering
\caption{Performance evaluation over comparison methods for Propagation Trees Identification (Best is highlighted with bold.)}
\label{tab:results_new}
\resizebox{\textwidth}{!}{%
\begin{tabular}{l@{\hspace{5pt}}cc@{\hspace{5pt}}cc@{\hspace{5pt}}cc@{\hspace{5pt}}cc@{\hspace{5pt}}cc}
\toprule
\multirow{2}{*}{Methods} & \multicolumn{2}{c}{Cora-ML} & \multicolumn{2}{c}{Memetracker} & \multicolumn{2}{c}{CiteSeer} & \multicolumn{2}{c}{Power Grid} & \multicolumn{2}{c}{Infection Spread Simulated Data} \\ 
\cmidrule(lr){2-3} \cmidrule(lr){4-5} \cmidrule(lr){6-7} \cmidrule(lr){8-9} \cmidrule(lr){10-11}
 & \small{Path Precision} & \small{Jaccard Index} & \small{Path Precision} & \small{Jaccard Index} & \small{Path Precision} & \small{Jaccard Index} & \small{Path Precision} & \small{Jaccard Index} & \small{Path Precision} & \small{Jaccard Index} \\ 
\midrule
DDMIX   & 0.327  & 0.195  & 0.062  & 0.041  & 0.236  & 0.133  & 0.081   & 0.031  & 0.109  & 0.057  \\
DDMSL   & 0.412  & 0.259  & 0.119  & 0.063   & 0.405  & 0.253  & 0.130   & 0.069  & 0.121  & 0.064  \\
DIPT & \textbf{0.622}  & \textbf{0.452}  & \textbf{0.602}  & \textbf{0.430}   & \textbf{0.593}  & \textbf{0.421}  & \textbf{0.680}   & \textbf{0.515}   & \textbf{0.421}  & \textbf{0.266}  \\
\bottomrule
\end{tabular}%
}
\end{table*}

\begin{table*}[h]
\centering
\caption{Performance evaluation over comparison methods for Source Localization (Best is highlighted with bold.)}
\label{tab:results}
\resizebox{\textwidth}{!}{%
\begin{tabular}{l@{\hspace{5pt}}cccc@{\hspace{5pt}}cccc@{\hspace{5pt}}cccc@{\hspace{5pt}}cccc@{\hspace{5pt}}cccc}
\toprule
\multirow{2}{*}{Methods} & \multicolumn{4}{c}{Cora-ML} & \multicolumn{4}{c}{Memetracker} & \multicolumn{4}{c}{CiteSeer} & \multicolumn{4}{c}{Power Grid} & \multicolumn{4}{c}{IDSS} \\ 
\cmidrule(lr){2-5} \cmidrule(lr){6-9} \cmidrule(lr){10-13} \cmidrule(lr){14-17} \cmidrule(lr){18-21}
 & \small{RE} & \small{PR} & \small{F1} & \small{AUC} & \small{RE} & \small{PR} & \small{F1} & \small{AUC} & \small{RE} & \small{PR} & \small{F1} & \small{AUC} & \small{RE} & \small{PR} & \small{F1} & \small{AUC} & \small{RE} & \small{PR} & \small{F1} & \small{AUC} \\ \midrule
LPSI  & 0.217 & 0.492 & 0.301 & 0.592 & 0.292 & 0.007 & 0.014 & 0.529 & 0.225 & 0.480 & 0.306 & 0.598 & 0.495 & 0.455 & 0.474 & 0.934 & 0.280 & 0.020 & 0.037 & 0.540 \\
OJC   & 0.119 & 0.123 & 0.121 & 0.534 & 0.022 & 0.031 & 0.026 & 0.517 & 0.115 & 0.118 & 0.117 & 0.530 & 0.287 & 0.104 & 0.153 & 0.501 & 0.025 & 0.033 & 0.028 & 0.520 \\
GCNSI  & 0.456 & 0.357 & 0.401 & 0.687 & 0.234 & 0.019 & 0.035 & 0.422 & 0.440 & 0.345 & 0.387 & 0.680 & 0.335 & 0.325 & 0.330 & 0.639 & 0.245 & 0.025 & 0.045 & 0.430 \\
SL-VAE & 0.719 & 0.814 & 0.764 & 0.831 & 0.518 & \textbf{0.461} & 0.488 & 0.624 & 0.700 & 0.805 & 0.749 & 0.825 & 0.780 & 0.815 & 0.797 & \textbf{0.879} & 0.520 & \textbf{0.470} & 0.494 & 0.630 \\
DDMIX  & 0.210 & 0.232 & 0.221 & 0.247 & 0.023 & 0.021 & 0.022 & 0.417 & 0.205 & 0.225 & 0.215 & 0.250 & 0.345 & 0.235 & 0.280 & 0.340 & 0.030 & 0.028 & 0.029 & 0.425 \\
DDMSL  & 0.758 & 0.742 & 0.750 & 0.873 & \textbf{0.618} & 0.441 & 0.515 & \textbf{0.641} & 0.750 & 0.735 & 0.742 & 0.870 & 0.763 & \textbf{0.913} & \textbf{0.831} & 0.866 & \textbf{0.625} & 0.455 & \textbf{0.527} & \textbf{0.645} \\
DIPT & \textbf{0.856} & \textbf{0.823} & \textbf{0.839} & \textbf{0.881} & 0.607 & 0.452 & \textbf{0.518} & 0.629 & \textbf{0.850} & \textbf{0.815} & \textbf{0.832} & \textbf{0.880} & \textbf{0.781} & 0.882 & 0.828 & 0.864 & 0.610 & 0.460 & 0.525 & 0.630 \\
\bottomrule
\end{tabular}%
}
\end{table*}
We illustrate the performance of DIPT on various experiments against two sets of methods.\newline
\begin{itemize}[left=0pt]
    \item\textit{Generative Diffusion Models.} These methods model the source localization problem using generative diffusion models and are capable of predicting node states at discrete timestep during the diffusion process. DDMIX~\cite{vcutura2021deep} employs a generative model for reconstructing information diffusion paths by learning node states across multiple time steps using a Variational Autoencoder (VAE). DDMSL~\cite{yan2024diffusion} uses a invertible diffusion-based generative model for source localization and recovering each diffusion step. However, neither method reconstructs the exact propagation tree (i.e., the edges through which information spreads). To the best of our knowledge, no existing work provides this information. Since these models predict node states at discrete timesteps during diffusion but not explicit edges, we adapt these models to reconstruct propagation path by combining random forward and backward walks among diffused states between two diffusion steps, establishing a baseline for comparison with our approach.\newline 
    \item \textit{Source Localization Methods.}  To evaluate DIPT's performance on the source localization task, we compare it with four baseline methods, in addition to DDMIX and DDMSL: 1). LPSI~\cite{wang2017multiple} predicts rumor sources based on the convergent node labels, without requiring knowledge of the underlying information propagation model. 2). OJC~\cite{zhu2017catch} specializes in source localization in networks with partial observations, demonstrating particular strength in detecting sources under the SIR diffusion pattern. 3). Among learning-based methods, GCNSI~\cite{dong2019multiple} learns latent node embeddings using a Graph Convolutional Network (GCN) to identify multiple rumor sources, closely matching the actual source. 4). SL-VAE~\cite{ling2022source} learn the graph diffusion model with a generative model to characterize the distribution of diffusion sources. 
\end{itemize}
\subsubsection{\textbf{Implementation Details:}}
We use a three-layer MLP project node features into a lower-dimensional space. A cross-attention module fuses the transformed features, followed by another 2-layer MLP to compute the influence score between two nodes. The attention mechanism follows the scaled dot-product formulation. Training is performed using the Adam optimizer with a learning rate of \( \eta \) for \( T \) epochs. For learning a prior over seed nodes, we employ a three-layer MLP with nonlinear transformations for both the encoder \( q_{\phi_1}(z|x) \) and decoder \( p_{\phi_2}(s|z) \). The learning rate is set to 0.005, with 500 training epochs for all datasets. Inference is performed with 100 iterations across all datasets.
\subsubsection{\textbf{Evaluation Metrics}.}
For evaluation, we use both propagation tree and source localization metrics to assess DIPT's performance. To measure the accuracy of the predicted propagation tree, we employ two metrics: the Jaccard Index and Path Precision. The Jaccard Index captures the overall overlap of edges between the predicted and actual propagation trees, providing a measure of general similarity, while Path Precision focuses on how accurately the model identifies the correct edges within the true propagation path. Together, these metrics assess both the completeness and accuracy of the reconstructed propagated path.

For source localization, which is a classification task, we evaluate performance using four primary metrics: Precision, Recall, F1-Score (F1), and ROC-AUC Curve (AUC). These are standard metrics for classification tasks and are commonly used in source localization studies.

\subsection{Propagation Trees Prediction Performance}

\begin{table*}[!h]
\centering
\caption{Performance of DIPT with varying proportions of partially observed propagation tree data during training }
\label{tab:partialObservation}
\resizebox{\textwidth}{!}{%
\begin{tabular}{c@{\hspace{5pt}}cccc@{\hspace{5pt}}cccc@{\hspace{5pt}}cccc@{\hspace{5pt}}cccc@{\hspace{5pt}}cccc}
\toprule
\multirow{2}{*}{Data Percentage} & \multicolumn{2}{c}{Cora-ML} & \multicolumn{2}{c}{Memetracker} & \multicolumn{2}{c}{CiteSeer} & \multicolumn{2}{c}{Power Grid} & \multicolumn{2}{c}{IDSS} \\ 
\cmidrule(lr){2-3} \cmidrule(lr){4-5} \cmidrule(lr){6-7} \cmidrule(lr){8-9} \cmidrule(lr){10-11}
 & \small{Path Precision} & \small{Jaccard Index} & \small{Path Precision} & \small{Jaccard Index} & \small{Path Precision} & \small{Jaccard Index} & \small{Path Precision} & \small{Jaccard Index} & \small{Path Precision} & \small{Jaccard Index} \\ \midrule
10\%  & 0.671 & 0.504 & 0.633 & 0.463 & 0.662 & 0.495 & 0.683 & 0.518 & 0.437 & 0.279 \\
20\%  & 0.707 & 0.546 & 0.651 & 0.482 & 0.671 & 0.504 & 0.718 & 0.560 & 0.451 & 0.291 \\
30\%  & 0.720 & 0.562 & 0.662 & 0.494 & 0.695 & 0.532 & 0.759 & 0.611 & 0.453 & 0.292 \\
\bottomrule
\end{tabular}%
}
\end{table*}

\begin{table*}[!h]
\centering
\caption{Performance of DIPT under Different Ablation Settings}
\label{tab:ablation_results}
\resizebox{\textwidth}{!}{%
\begin{tabular}{c@{\hspace{5pt}}cccc@{\hspace{5pt}}cccc@{\hspace{5pt}}cccc@{\hspace{5pt}}cccc@{\hspace{5pt}}cccc}
\toprule
\multirow{2}{*}{Ablation Setting} & \multicolumn{2}{c}{Cora-ML} & \multicolumn{2}{c}{Memetracker} & \multicolumn{2}{c}{CiteSeer} & \multicolumn{2}{c}{Power Grid} & \multicolumn{2}{c}{IDSS} \\ 
\cmidrule(lr){2-3} \cmidrule(lr){4-5} \cmidrule(lr){6-7} \cmidrule(lr){8-9} \cmidrule(lr){10-11}
 & \small{Path Precision} & \small{Jaccard Index} & \small{Path Precision} & \small{Jaccard Index} & \small{Path Precision} & \small{Jaccard Index} & \small{Path Precision} & \small{Jaccard Index} & \small{Path Precision} & \small{Jaccard Index} \\ \midrule
\textsc{DIPT (a)}  & 0.388 & 0.235 & 0.407 & 0.255 & 0.422 & 0.267 & 0.437 & 0.276 & 0.329 & 0.206 \\ 
{DIPT (b)}  & 0.519 & 0.350 & 0.489 & 0.326 & 0.511 & 0.343 & 0.607 & 0.435 & 0.371 & 0.227 \\
{DIPT}  & 0.622  & 0.452  & 0.602  & 0.430   & 0.593  & 0.421  & 0.680   & 0.515   & 0.421  & 0.266  \\  
\bottomrule
\end{tabular}%
}
\end{table*}
The performance of the proposed model, DIPT, compared to two baseline approaches for propagation tree identification is summarized in Table~1. Since DDMIX and DDMSL cannot directly infer propagation trees in their default settings but instead recover propagation snapshots (node states) at discrete timesteps, we adapt them as described in Section~\ref{sec:comparMethod} for direct comparison. DIPT consistently outperforms both methods across all five datasets on both propagation tree evaluation metrics.  

The performance gap is particularly pronounced in the denser datasets—Power Grid, IDSS, and MemeTracker—where existing approaches struggle due to increased graph density and complex propagation patterns. In contrast, on the sparser CiteSeer and Cora-ML datasets, the difference is less noticeable. This highlights DIPT’s ability to accurately infer propagation trees even in dense graphs, as it learns the influence of one node over another, as shown in~\eqref{eq:07}. DIPT achieves its best performance on Power Grid, with 68\% path precision and a 51.5\% Jaccard index. IDSS is the most challenging dataset due to high variability in inter-county mobility and outbreak patterns; yet, DIPT attains a path precision of 42.1\%, far exceeding DDMSL (6.9\%) and DDMIX (3.1\%).  

Figure~\ref{fig:comparison_memetracker} compares propagation trees identified by all methods to the ground truth on memetracker dataset. Since each infected node has exactly one infecting parent, all methods predict the same number of edges, and for clarity, only true positives(blue directed edges) are shown. The figure confirms that in denser regions (light gray lines indicate edges), DDMIX and DDMSL’s accuracy drops significantly, whereas DIPT performs well due to its framework’s ability to learn local influence scores. Additional visualizations are shown in Appendix.  
On average, DIPT achieves a 3.5x and 4.37× relative gain in path precision and Jaccard index, respectively, compared to DDMSL and DDMIX across all datasets.  
\subsection{Source Localization Accuracy}
We evaluate DIPT against other source localization methods using the SI epidemic model as the underlying propagation model. For comparison, we use the six methods discussed in Section~\ref{sec:comparMethod}, and present the performance results in Table~2. As shown, DIPT performs competitively with DDMSL and SL-VAE across all datasets, achieving notable improvements over the other source localization methods. Due to the highly imbalanced ratio of diffusion sources to other nodes, methods like LPSI, OJC, and GCNSI struggle to capture the diffusion sources' distribution, resulting in inaccurate predictions of the number of source nodes.
DIPT achieves the best performance on the Cora-ML and CiteSeer datasets, with F1 and AUC scores of 0.839 and 0.881, respectively, for Cora-ML. This aligns with DIPT’s performance in predicting propagation tree structures, as these datasets are relatively less dense compared to others. On the IDSS dataset, where the distribution of seed nodes varies due to mobility patterns from airports, DIPT performs similarly to DDMSL, with F1 and AUC scores of 0.524 and 0.630, respectively.
The results in Table~2 indicate that the proposed model design effectively preserves source localization performance, as source localization is a subset of propagation tree identification (i.e., identifying the roots of the propagation trees).
\subsection{Diffusion State Reconstruction}
\begin{figure}[t] 
    \centering
    \includegraphics[width=\columnwidth]{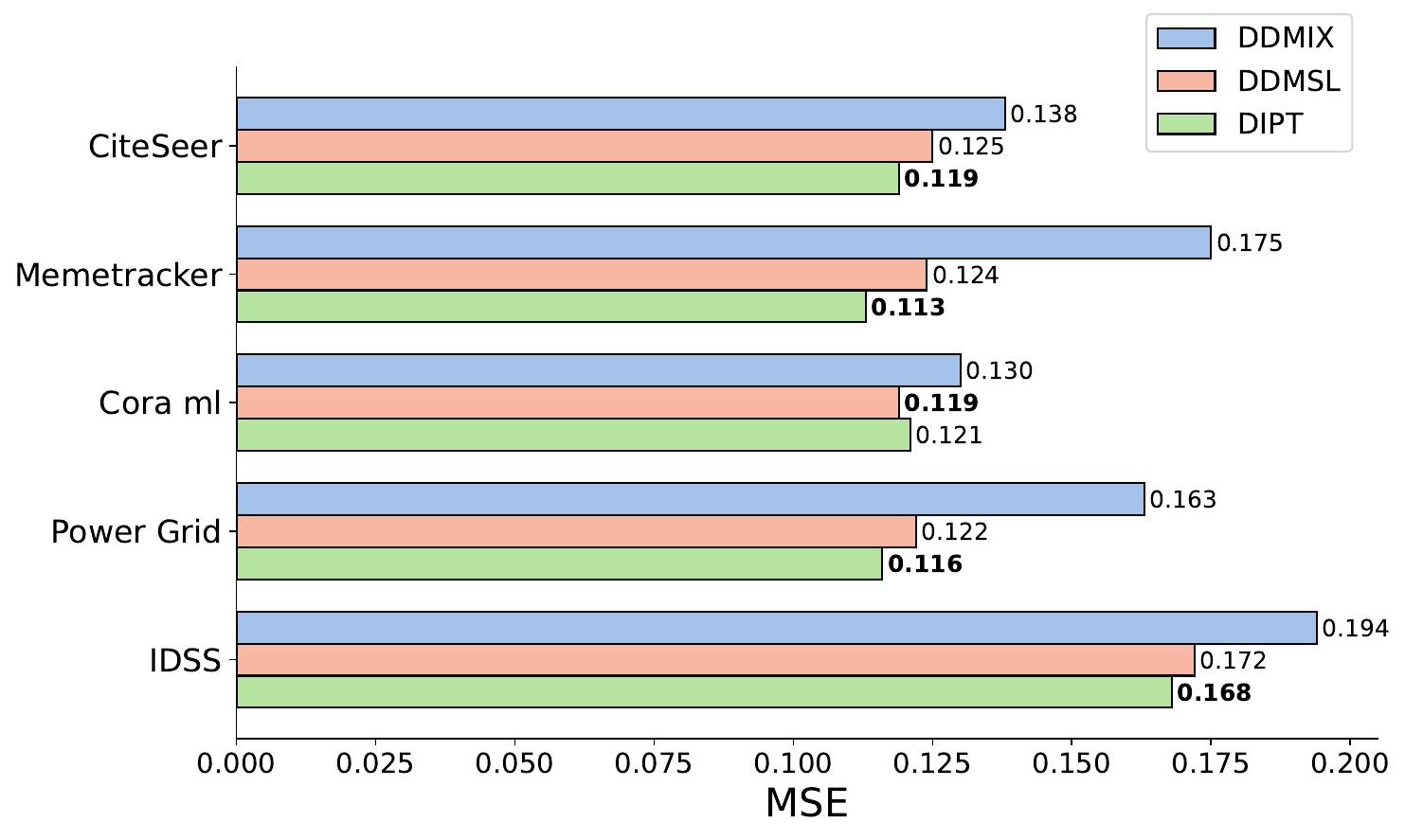} %
    \caption{Comparison of DIPT, DDMSL and DDMIX on reconstructed information diffusion process}
    \label{fig:MSE}
\end{figure}

Both DDMSL and DDMIX reconstruct node states during the diffusion process, providing information on which nodes are infected at each discrete time step, resulting in a sequence of infections that shows the order in which nodes are infected. DDMIX only recovers the states of susceptible (S) and infected (I) nodes, while DDMSL can reconstruct all possible node states. In contrast, the proposed model, DIPT, not only recovers the infection sequence (i.e., the order in which nodes are infected) but also identifies the source of each infection (i.e., who infected whom), effectively reconstructing the entire propagation tree.
To make a fair comparison, we evaluate node state reconstruction based on the infection sequence during the diffusion process, with results summarized in Fig.~\ref{fig:MSE}. DIPT achieves an average mean squared reconstruction error that is 19.67\% and 3.85\% lower than DDMIX and DDMSL, respectively. These results demonstrate that, in addition to excelling in propagation tree identification, DIPT accurately reconstructs node states during the diffusion process.

\subsection{Impact of Partial Observations}
Since the proposed method is primarily designed to learn propagation trees without access to propagation tree data during training, Section~\ref{sec:partial} demonstrates how partial information about propagation trees can be incorporated. In Table~\ref{tab:partialObservation}, we present the impact on DIPT’s performance when 10\%, 20\%, and 30\% of propagation tree information is available during training. With these partial observations, DIPT achieves a 7.11\%, 11.84\%, and 15.16\% increase in path precision, respectively, on average across all datasets.

The observed performance improvement with partial data can be explained by the fact that directly supervising the available partial propagation information, as shown in Eq.~\ref{eq:supervisedEdge}, helps the model better learn local influences between nodes. However, during inference, edges are selected based on their ability to maximize the likelihood of the observed infection. As a result, the performance gains are not directly proportional to the percentage of available data, which explains why the improvement is not linear with the percentage of partial observations used.

\subsection{Ablation Results}
We conducted an ablation study to evaluate the contribution of each component in DIPT. In the first ablated model (DIPT(a)), instead of learning the influence between two nodes as described in Eq.\ref{eq:07}, we approximate it using the cosine similarity between the features of the nodes at both ends of the edge. In the second ablated model (DIPT(b)), we directly optimize the joint objective in Eq.\ref{eq:objective} without alternately inferring the propagation tree and optimizing the objective function during training. However, during evaluation at inference, the propagation tree is inferred as described in Section~\ref{sec:propTreeInference}. The results are summarized in Table~\ref{tab:ablation_results}.

Removing any component from DIPT leads to a significant reduction in performance. In DIPT(a), the model's path accuracy drops sharply because it no longer learns the local node influence across edges. Without this learning module, the influence scores remain constant, severely affecting performance in propagation tree identification. On the Cora-ML dataset, DIPT(a) even performs worse than DDMSL in path precision (Table~\ref{tab:results_new}). However, for denser graphs, it still outperforms other comparison methods from Table~\ref{tab:results_new} in identifying propagation paths. DIPT(b) also shows a performance decline, but it is less severe than DIPT(a). Despite this, DIPT(b) still outperforms all comparison methods across datasets, demonstrating that learning local influence between nodes is more crucial for capturing diffusion patterns. The performance gap between DIPT and DIPT(b) emphasizes the importance of alternately updating the most probable propagation tree during training.

\section{Conclusion}
Identification of propagation trees is a crucial yet underexplored task with significant applications in fields such as epidemiology and misinformation diffusion. In this paper, we introduce DIPT, a probabilistic framework designed to identify propagation trees from observed diffusion data. DIPT recursively models the diffusion process by learning influence probabilities across edges, informed by node features. The framework employs an alternating optimization approach to jointly learn both the propagation tree and the diffusion mechanism, without relying on direct observation of propagation paths during training. Extensive experiments on five datasets demonstrate that DIPT consistently outperforms existing methods, achieving an average path precision of 58.2\% and effectively identifying both propagation trees and diffusion sources.
\bibliographystyle{ACM-Reference-Format}
\bibliography{reference}
\appendix
\begin{figure*}[h] 
    \centering
    \includegraphics[width=1.00\linewidth]{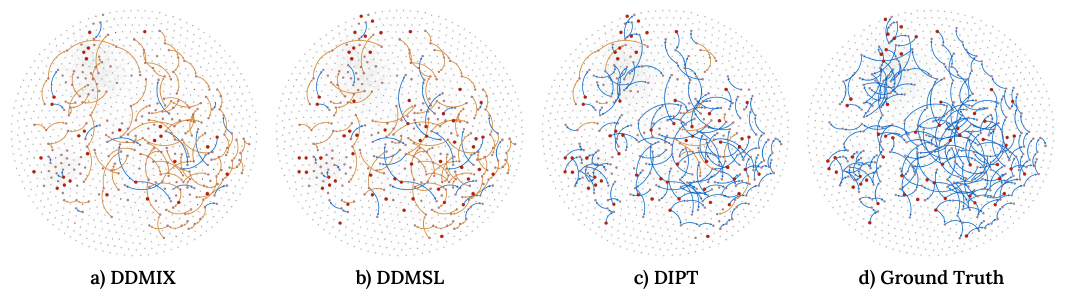}
    \caption{Comparison of predicted propagation tree edges with ground truth for the MemeTracker dataset. Source nodes are in red, infected nodes in pink.}
\end{figure*}

\begin{figure*}[h] 
    \centering
    \includegraphics[width=1.00\linewidth]{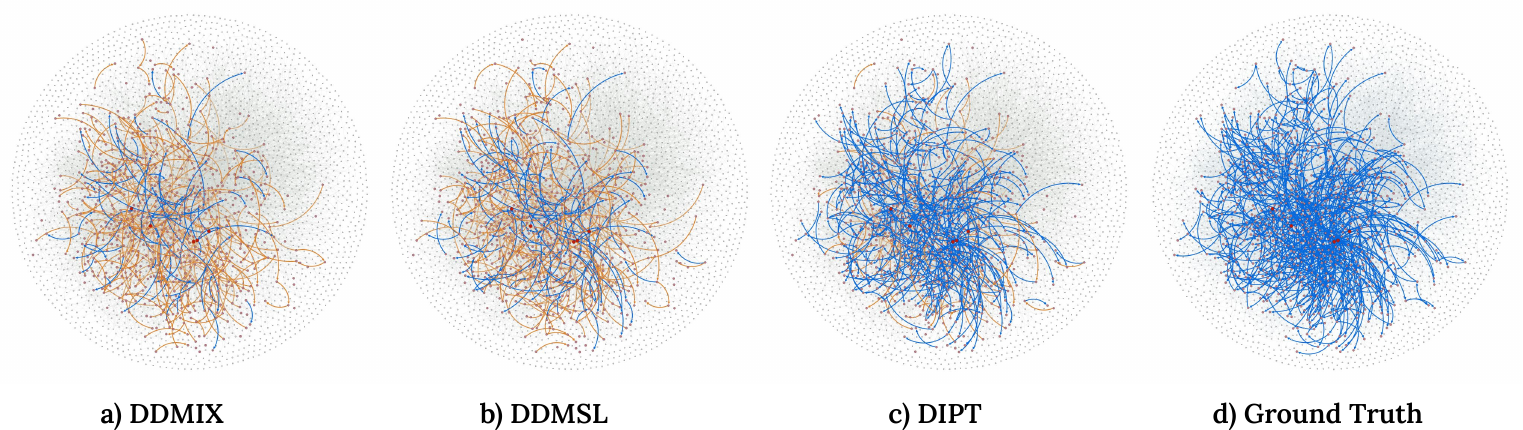} %
    \caption{Comparison of predicted propagation tree edges with ground truth for the IDSS dataset. Source nodes are in red, infected nodes in pink.}
\end{figure*}
\section{Datasets Description}
We conduct experiments on four real-world datasets and one simulated dataset (explained in Section~\ref{sec:Simulated}):

\begin{itemize}
    \item \textbf{Cora-ML}~\cite{cora}. This network contains computer science research papers, where each node represents a paper and each edge indicates that one paper cites another.
    
    \item \textbf{Power Grid}~\cite{watts1998collective}. This is the topology network of the Western States Power Grid of the US. An edge represents a power supply line, and a node is either a generator, a transformer, or a substation.

    \item \textbf{Memetracker}~\cite{leskovec2009meme}. MemeTracker tracks the posts that appear most frequently over time across the entire online news spectrum. The propagation of each story is represented as one diffusion cascade.

\item \textbf{CiteSeer}~\cite{caragea2014citeseer}. This is a citation network of research papers, where each node represents a paper, and edges indicate citation relationships. Papers are classified into different categories based on their research topics.

\end{itemize}
\section{Additional Visualizations}
We provide additional visualizations of propagation tree identification for the MemeTracker and simulated IDSS dataset. Blue edges represent correctly identified propagation edges, while orange edges indicate incorrectly predicted ones.



\end{document}